\documentclass[12pt]{article}
\usepackage{amsmath,amssymb,bm,graphicx}
\usepackage{color} 
\usepackage{hyperref}

\newcommand{\pslash}{\not \! p}

\newcommand{\delslash}{\not \! \partial}

\usepackage{cite}

\begin{document}

%\vskip 0.5 truecm

\begin{center}
{\Large{\bf Majorana neutrinos in seesaw mechanism\\ and Bogoliubov quasiparticles }}
\end{center}
\vskip .5 truecm
\begin{center}
{\bf { Kazuo Fujikawa$^{1,2}$ and Anca Tureanu$^1$}}
\end{center}

\begin{center}
\vspace*{0.4cm} 
{\it {$^1$Department of Physics, University of Helsinki, P.O.Box 64, 
\\FIN-00014 Helsinki,
Finland\\
$^2$Quantum Hadron Physics Laboratory, RIKEN Nishina Center,\\
Wako 351-0198, Japan
}}
\end{center}
\begin{abstract}
The idea that the Majorana neutrino should be identified as a Bogoliubov quasiparticle is applied to the seesaw mechanism for the three generations of neutrinos in the Standard Model. A relativistic analogue of  the Bogoliubov transformation in the present context is a  CP-preserving canonical transformation but modifies  charge conjugation properties in such a way that the C-noninvariant fermion number violating term (condensate) is converted to a Dirac mass term.  Puzzling aspects associated with the charge conjugation of chiral Weyl fermions are clarified.   By invoking the Coleman--Weinberg mechanism in the framework of dimensional regularization, it is also noted that seesaw models become unnatural in some parameter regions which induce the hierarchy problems in the bosonic sector.
\end{abstract}
%\maketitle
%\large
\section{Introduction}

We have recently witnessed a remarkable progress in neutrino 
physics~\cite{particledata}. Those achievements in experimental and theoretical studies of  neutrino physics are nicely summarized in many textbooks and reviews, for example, ~\cite{fukugita, giunti, bilenky, xing,valle}. The major remaining issue is a better understanding of the extremely small neutrino masses, and the seesaw mechanism provides a convenient framework to analyze this fundamental problem~\cite{minkowski,yanagida,mohapatra}.  Recently we discussed the basic issue of Majorana neutrinos  in the seesaw mechanism~\cite{FT,FT2}. A relativistic analogue of the Bogoliubov transformation, which converts a C-noninvariant fermion number ``condensate'' to a Dirac mass and changes the charge conjugation properties of vacuum, is shown to be crucial to understand the Majorana neutrinos in a logically consistent manner. 

To be more explicit, a relativistic analogue of the  Bogoliubov transformation has been introduced to resolve the well-known puzzling feature of the charge conjugation of chiral Weyl fermions
\begin{eqnarray}
\nu_{L}(x)=\frac{(1-\gamma_{5})}{2}\nu(x).
\end{eqnarray}
The commonly adopted convention of charge conjugation
\begin{eqnarray}\label{naive-C}
(\nu_{L}(x))^{C}\equiv C\overline{\nu_{L}}^{T}(x),
\end{eqnarray}
which is required to define the Majorana fermion in the seesaw mechanism, changes at the same time the chirality (helicity) of the neutrino  ~\cite{fukugita, giunti, bilenky, xing}. This definition leads to many puzzling results~\cite{FT}. Those contradictions are resolved if one uses a relativistic analogue of  Bogoliubov transformation, which is a canonical transformation and converts a C-noninvariant fermion number ``condensate'' to a Dirac mass, and we suggested the idea that the Majorana neutrino should be identified as the first Bogoliubov quasiparticle among elementary  particles~\cite{FT2}. We emphasize that the above-mentioned puzzling features turned out to be not a matter of notational convention but rooted in an important conceptual issue of the fermion vacuum.

In this paper, we extend the idea of the Majorana neutrino being the first Bogoliubov quasiparticle among elementary  particles to the full three generations of neutrinos. 
We study the Lagrangian for the three families of neutrinos,
\begin{eqnarray}\label{1}
{\cal L}&=&\overline{\nu}_{L}(x)i\gamma^{\mu}\partial_{\mu}\nu_{L}(x)+\overline{n}_{R}(x)i\gamma^{\mu}\partial_{\mu}n_{R}(x)\nonumber\\
&-&\overline{\nu}_{L}(x)m_{D} n_{R}(x)
-(1/2)\nu_{L}^{T}(x)C m_{L}\nu_{L}(x)\nonumber\\
&-&(1/2)n_{R}^{T}(x)C m_{R}n_{R}(x) + h.c.,
\end{eqnarray}
where $n_{R}(x)$ is a right-handed counterpart of $\nu_{L}(x)$; $m_{D}$ is a diagonal $3\times 3$ Dirac mass matrix, and $m_{L}$ and $m_{R}$ are $3\times 3$ real symmetric matrices by assuming CP symmetry, for simplicity. The anti-symmetry of $C$ and Fermi statistics imply that $m_{L}$ and $m_{R}$ are symmetric, and  CP symmetry implies 
$m_{L}=m_{L}^{\dagger}$ and $m_{R}=m_{R}^{\dagger}$. Thus $m_{L}$ and $m_{R}$ are real symmetric.
We follow the notational conventions of \cite{bjorken}.

One may define a new Dirac-type variable
\begin{eqnarray}\label{2}
\nu(x)\equiv \nu_{L}(x) + n_{R}(x),
\end{eqnarray}
in terms of which the above Lagrangian is re-written as
\begin{eqnarray}\label{3}
{\cal L}&=&(1/2)\{\overline{\nu}(x)[i\delslash - m_{D}]\nu(x)+ \overline{\nu^{c}}(x)[i\delslash - m_{D}]\nu^{c}(x)\}\nonumber\\
&-&(1/4)[\overline{\nu^{c}}(x)\epsilon_{1}\nu(x) +\overline{\nu}(x)\epsilon_{1}\nu^{c}(x)]\nonumber\\
&-&(1/4)[\overline{\nu^{c}}(x)\gamma_{5}\epsilon_{5}\nu(x) -\overline{\nu}(x)\gamma_{5}\epsilon_{5}\nu^{c}(x)],
\end{eqnarray}
where $\epsilon_{1}=m_{R}+m_{L}$ and $\epsilon_{5}=m_{R}-m_{L}$, which are real symmetric matrices. 
The C and P transformation rules for the Dirac-type field $\nu(x)$ are defined by 
\begin{eqnarray}\label{4}
\nu^{c}(x)=C\bar{\nu}^{T}(x), \ \ \nu^{p}(x)=i\gamma^{0}\nu(t,-\vec{x}), 
\end{eqnarray}
and thus $\nu(x) \leftrightarrow \nu^{c}(x)$ under C and $\nu^{c}(x)\rightarrow i\gamma^{0}\nu^{c}(t,-\vec{x})$ under P; we adopt the charge conjugation convention in which $C=i\gamma^{2}\gamma^{0}$ and the "$i\gamma^{0}$-parity", $\nu^{p}(x)=i\gamma^{0}\nu(t,-\vec{x})$ as in \eqref{4}, since they preserve the reality of the Majorana fermion in the Majorana representation.
CP is then given by 
\begin{eqnarray} \label{CP}
\nu^{cp}(x)=i\gamma^{0}C\bar{\nu}^{T}(t,-\vec{x}).
\end{eqnarray}
The above  Lagrangian \eqref{3} is CP conserving, although C and P ($i\gamma^{0}$-parity) are separately broken by the last term. We
are going to explain how the C-violating Lagrangian \eqref{3}
can consistently describe Majorana neutrinos which are the
exact eigenstates of C symmetry.

\section{Seesaw mechanism}

\subsection{Perturbative formulation}

To set up the stage for the analysis of the Bogoliubov quasi-fermion in the seesaw mechanism, we first consider the simplest case with $m_{L}=0$ and treat the Dirac mass term containing $m_{D}$ in \eqref{3} as a small perturbation.
For this purpose, we apply an orthogonal transformation in \eqref{3}, $\nu(x)\rightarrow O\nu(x)$ (and thus $\nu^{c}(x)\rightarrow O\nu^{c}(x)$), and diagonalize the real symmetric matrix $\epsilon_{1}=\epsilon_{5}=m_{R}$; after this operation the Dirac mass $m_{D}$ becomes a $3\times 3$ non-diagonal real symmetric matrix. We thus have the zeroth order Lagrangian:
\begin{eqnarray}\label{free-Lagrangian_1}
{\cal L}_{0}&=&(1/2)\{\overline{\nu}(x)i\delslash \nu(x)+ \overline{\nu^{c}}(x)i\delslash \nu^{c}(x)\}\nonumber\\
&-&(1/4)[\overline{\nu^{c}}(x)m_{R}\nu(x) +\overline{\nu}(x)m_{R}\nu^{c}(x)]\nonumber\\
&-&(1/4)[\overline{\nu^{c}}(x)\gamma_{5}m_{R}\nu(x) -\overline{\nu}(x)\gamma_{5}m_{R}\nu^{c}(x)].
\end{eqnarray}
This Lagrangian is diagonalized as 
\begin{eqnarray}\label{5}
{\cal L}_{0}
&=&(1/2)\overline{\psi_{-}}(x)i\gamma^{\mu}\partial_{\mu}\psi_{-}(x)\nonumber\\
&+&(1/2)\overline{\psi_{+}}(x)i\gamma^{\mu}\partial_{\mu}\psi_{+}(x)
-(1/2)\overline{\psi_{+}}(x) m_{R}\psi_{+}(x)
\end{eqnarray}
if one parameterizes
\begin{eqnarray}\label{free-neutrino}
            \left(\begin{array}{c}
            \nu(x)\\
            \nu^{c}(x)
            \end{array}\right)
            &=&\left(\begin{array}{c}
            \frac{1+\gamma_{5}}{2}\psi_{+}(x)+\frac{1-\gamma_{5}}{2}      
            \psi_{-}(x)\\
            \frac{1-\gamma_{5}}{2}\psi_{+}(x)-\frac{1+\gamma_{5}}{2}      
            \psi_{-}(x)            
            \end{array}\right),
\end{eqnarray}
or equivalently,
\begin{eqnarray}
\psi_{+}(x)&=&\nu_{R}(x)+\nu_{L}^{c}(x),\nonumber\\
\psi_{-}(x)&=&\nu_{L}(x)-\nu_{R}^{c}(x).
\end{eqnarray}
Note that  $\nu_{L}^{c}\equiv [(1-\gamma_{5})/2]\nu^{c}$ and $\nu_{R}^{c}\equiv [(1+\gamma_{5})/2]\nu^{c}$, and they are left- and right-handed, respectively. Our {\em convention} of charge conjugation for chiral fermions, which turns out to be convenient in the following analyses, thus differs from the convention commonly used in the seesaw mechanism in \eqref{naive-C}.

The C-symmetry $\nu^{c}(x)=C\overline{\nu(x)}^{T}$ is satisfied in \eqref{free-neutrino} if the following relations hold:
\begin{eqnarray}\label{7}
\psi^{c}_{+}(x)&=&C\overline{\psi_{+}(x)}^{T}=\psi_{+}(x),\nonumber\\
\psi^{c}_{-}(x)&=&C\overline{\psi_{-}(x)}^{T}=-\psi_{-}(x),
\end{eqnarray}
showing that the free fields $\psi_{+}(x)$ and $\psi_{-}(x)$ are Majorana fermions.
The small Dirac mass term is written as 
\begin{eqnarray}\label{Dirac-mass}
{\cal L}_{mass}&=&-\frac{1}{2}[\overline{\nu}(x)m_{D} \nu(x)+ 
\overline{\nu^{c}}(x)m_{D} \nu^{c}(x)]\nonumber\\
&=&\frac{1}{2}[\overline{\psi}_{+}\gamma_{5}m_{D}\psi_{-}-\overline{\psi}_{-}\gamma_{5}m_{D} \psi_{+}].
\end{eqnarray}
One may perform a second-order perturbation analysis by treating
the Dirac mass term $m_{D}$ in \eqref{3} as a small perturbation.
The result is, symbolically,
\begin{eqnarray}\label{9}
 (1/2!)\overline{\psi}_{-}\gamma_{5}m_{D}\langle T\psi_{+}\overline{\psi}_{+}\rangle m_{D}\gamma_{5} \psi_{-}
\simeq (-i/2)\overline{\psi}_{-}m_{D}\frac{1}{m_{R}}m_{D}
\psi_{-},
\end{eqnarray}
using $\langle T\psi_{+}\overline{\psi}_{+}\rangle=\frac{i}{\pslash-m_{R}}$ near {\em on-shell} $\pslash=0$ of $\psi_{-}$. A quick way to obtain the above result is to exponentiate ${\cal L}_{mass}$ in the Dyson formula and expand it to the second order in ${\cal L}_{mass}$; by contracting the fields $\psi_{+}$, one identifies the induced mass \eqref{9}. This symmetric mass term,
\begin{eqnarray}\label{10}
\tilde{m}_{ab}\equiv \sum_{c}(m_{D})_{ac}\frac{1}{m^{c}_{R}}(m_{D})_{cb},
\end{eqnarray}
is added to the massless fermion $\psi_{-}$ in \eqref{4}. 
We thus obtain the effective Lagrangian
\begin{eqnarray}\label{11}
{\cal L}_{{\rm effective}}
&=&(1/2)\overline{\psi_{-}}(x)i\gamma^{\mu}\partial_{\mu}\psi_{-}(x)-(1/2)\overline{\psi_{-}}(x) \tilde{m} \psi_{-}(x)\nonumber\\
&+&(1/2)\overline{\psi_{+}}(x)i\gamma^{\mu}\partial_{\mu}\psi_{+}(x)
-(1/2)\overline{\psi_{+}}(x) m_{R}\psi_{+}(x),
\end{eqnarray}
which shows that the massless neutrinos acquire small masses after the diagonalization of $\tilde{m}$.
This perturbative formulation is convenient to see the essence of  the {\em seesaw mechanism}. 

\subsection{Exact treatment}

To perform a general analysis of a relativistic analogue of the Bogoliubov transformation, an exact treatment of the seesaw Lagrangian is important.
We start with the Lagrangian \eqref{3} and write the mass term as 
\begin{eqnarray}
(-2){\cal L}_{mass}=
\left(\begin{array}{cc}
            \overline{\nu_{R}}&\overline{\nu_{R}^{c}}
            \end{array}\right)
\left(\begin{array}{cc}
            \frac{1}{2}(\epsilon_{1}+\epsilon_{5})& m_{D}\\
            m_{D}&\frac{1}{2}(\epsilon_{1}-\epsilon_{5})
            \end{array}\right)
            \left(\begin{array}{c}
            \nu_{L}^{c}\\
            \nu_{L}
            \end{array}\right) +h.c.,
\end{eqnarray}
where 
\begin{eqnarray}
\nu_{L}^{c}\equiv C\overline{\nu_{R}}^T, \ \ \ \nu_{R}^{c}\equiv C\overline{\nu_{L}}^T.  
\end{eqnarray}
Note again that these conventions {\em differ} from the quantity defined in the common convention in seesaw mechanism \eqref{naive-C}; in our definition, $\nu_{L}^{c}=[(1-\gamma_{5})/2]\nu^{c}$ and $\nu_{R}^{c}=[(1+\gamma_{5})/2]\nu^{c}$ are left- and right-handed, respectively. 
Since the mass matrix appearing is real and symmetric, we can diagonalize it 
by an orthogonal transformation as,
\begin{eqnarray}
            O
            \left(\begin{array}{cc}
            \frac{1}{2}(\epsilon_{1}+\epsilon_{5})& m_{D}\\
            m_{D}&\frac{1}{2}(\epsilon_{1}-\epsilon_{5})
            \end{array}\right)
            O^{T}
            =\left(\begin{array}{cc}
            M_{1}&0\\
            0&-M_{2}
            \end{array}\right)    ,        
\end{eqnarray}
where  $M_{1}$ and $M_{2}$ are $3\times 3$ real diagonal matrices. We denote one of the eigenvalues as $-M_{2}$ instead of $M_{2}$ to define the natural Majorana mass later.
We thus have
\begin{eqnarray}
(-2){\cal L}_{mass}
&=& \left(\begin{array}{cc}
            \overline{\tilde{\nu}_{R}}&\overline{\tilde{\nu}_{R}^{c}}
            \end{array}\right)
\left(\begin{array}{cc}
            M_{1}&0\\
            0&-M_{2} 
            \end{array}\right)            
            \left(\begin{array}{c}
            \tilde{\nu}_{L}^{c}\\
            \tilde{\nu}_{L}
            \end{array}\right) +h.c.,                       
\end{eqnarray}
with
\begin{eqnarray} \label{variable-change}          
            &&\left(\begin{array}{c}
            \tilde{\nu}_{L}^{c}\\
            \tilde{\nu}_{L}
            \end{array}\right)\equiv O            
            \left(\begin{array}{c}
            \nu_{L}^{c}\\
            \nu_{L}
            \end{array}\right)
            ,\nonumber\\ 
            &&\left(\begin{array}{c}
            \tilde{\nu}_{R}\\
            \tilde{\nu}_{R}^{c}
            \end{array}\right)\equiv O            
            \left(\begin{array}{c}
            \nu_{R}\\
            \nu_{R}^{c}
            \end{array}\right).          
\end{eqnarray}
Hence we can write
\begin{eqnarray}\label{exact-solution}
{\cal L}
&=&(1/2)\{\overline{\tilde{\nu}_{L}}(x)i\delslash \tilde{\nu}_{L}(x)+ \overline{\tilde{\nu}_{L}^{c}}(x)i\delslash \tilde{\nu}_{L}^{c}(x)+\overline{\tilde{\nu}_{R}}(x)i\delslash \tilde{\nu}_{R}(x)\nonumber\\
&&\ \ \ \ \ + \overline{\tilde{\nu}_{R}^{c}}(x)i\delslash \tilde{\nu}_{R}^{c}(x)\}\nonumber\\
&-&(1/2)\left(\begin{array}{cc}
            \overline{\tilde{\nu}_{R}}&\overline{\tilde{\nu}_{R}^{c}}
            \end{array}\right)
\left(\begin{array}{cc}
            M_{1}&0\\
            0&-M_{2} 
            \end{array}\right)            
            \left(\begin{array}{c}
            \tilde{\nu}_{L}^{c}\\
            \tilde{\nu}_{L}
            \end{array}\right) +h.c.
\end{eqnarray}
We now define two fields by
\begin{eqnarray}\label{Majorana}
\psi_{+}(x)&=&\tilde{\nu}_{R}(x)+\tilde{\nu}_{L}^{c}(x),\nonumber\\
\psi_{-}(x)&=&\tilde{\nu}_{L}(x)-\tilde{\nu}_{R}^{c}(x),
\end{eqnarray}
which satisfy the charge conjugation properties $\psi_{+}(x)=C\overline{\psi_{+}}^{T}(x)$ and $\psi_{-}(x)=-C\overline{\psi_{-}}^{T}(x)$,
remembering our convention $\tilde{\nu}_{L}^{c}=[(1-\gamma_{5})/2]\tilde{\nu}^{c}$ and $\tilde{\nu}_{R}^{c}=[(1+\gamma_{5})/2]\tilde{\nu}^{c}$. Equivalently, one can also write
\begin{eqnarray}\label{6}
            \left(\begin{array}{c}
            \tilde{\nu}(x)\\
            \tilde{\nu}^{c}(x)
            \end{array}\right)
            &=&\left(\begin{array}{c}
            \frac{1+\gamma_{5}}{2}\psi_{+}(x)+\frac{1-\gamma_{5}}{2}      
            \psi_{-}(x)\\
            \frac{1-\gamma_{5}}{2}\psi_{+}(x)-\frac{1+\gamma_{5}}{2}      
            \psi_{-}(x)            
            \end{array}\right).
\end{eqnarray}
We then have
\begin{eqnarray}\label{exact-solution2}
 {\cal L}
&=&\frac{1}{2}\overline{\psi_{+}}(x)[i\delslash - M_{1}]\psi_{+}(x)
+\frac{1}{2}\overline{\psi_{-}}(x)[i\delslash - M_{2}]\psi_{-}(x),            
\end{eqnarray}
which is the result of the {\em conventional} diagonalization~\cite{fukugita, giunti, bilenky, xing,valle}.

\section{Bogoliubov quasiparticle}

We now analyze the conflicts of the above seesaw mechanism with the operator charge conjugation symmetry. 
In quantum field theory the simple matrix operation \eqref{7} has to correspond to the application of a unitary $\cal C$ operator to the quantum fields. In the quantum framework, the definition of a charge conjugated spinor as $\psi^c = C\overline{\psi}^{T}$ can be regarded as a classical operation, for which a quantum realization $\cal C$ has to exist.

To simplify the notations and to discuss both the perturbative and the exact treatment  in the previous section on equal footing, we denote the fields $\nu(x)$ and $\nu^{c}(x)$, respectively, in this and next section standing for either $\nu(x)$ and $\nu^{c}(x)$ of the free Lagrangian in \eqref{free-Lagrangian_1}, or $\tilde{\nu}(x)$ and $\tilde{\nu}^{c}(x)$ of the exact Lagrangian in \eqref{exact-solution}.
%Note that the orthogonal transformation in \eqref{variable-change} faithfully transcribes C and P ($i\gamma^{0}$-parity) %transformation properties of $\nu(x)$ and $\nu^{c}(x)$  to $\tilde{\nu}(x)$ and $\tilde{\nu}^{c}(x)$, respectively.

When one defines the operator charge conjugation of Majorana fermions by 
\begin{eqnarray}\label{operator-C}
{\cal C}\psi_{+}(x){\cal C}^{\dagger}&=&\psi_{+}(x),\nonumber\\
{\cal C}\psi_{-}(x){\cal C}^{\dagger}&=&-\psi_{-}(x),
\end{eqnarray}
one can confirm that this operation in \eqref{6} does not send $\nu(x)$ to $\nu^{c}(x)$, although the classical C symmetry $\nu^{c}(x)=C\overline{\nu(x)}^{T}$ is satisfied.  

This conflict is also understood by analyzing \eqref{Majorana}, which is re-written below in the notation of the present section as
\begin{eqnarray}
\psi_{+}(x)&=&\nu_{R}(x)+\nu_{L}^{c}(x),\nonumber\\
\psi_{-}(x)&=&\nu_{L}(x)-\nu_{R}^{c}(x).\nonumber
\end{eqnarray}
To ensure the consistent operator charge conjugation of $\psi_{\pm}(x)$ in the above formulas, recalling that $\nu_{L}^{c}=[(1-\gamma_{5})/2]\nu^{c}=C\overline{\nu_{R}}^{T}$ and $\nu_{R}^{c}=[(1+\gamma_{5})/2]\nu^{c}=C\overline{\nu_{L}}^{T}$, it is commonly assumed~\cite{fukugita, giunti, bilenky, xing} that one can find  a suitable charge conjugation operator ${\cal C}$ (on an unspecified vacuum) which satisfies 
\begin{eqnarray}\label{naive-operator-C}
{\cal C}\nu_{R}(x){\cal C}^{\dagger}=C\overline{\nu_{R}}^{T},\ \ \
{\cal C}\nu_{L}(x){\cal C}^{\dagger}=C\overline{\nu_{L}}^{T},
\end{eqnarray}
and thus \eqref{operator-C} is satisfied.
But this leads to a puzzling result for the unitary charge conjugation operator using $\nu_{L}(x)=\frac{(1-\gamma_{5})}{2}\nu_{L}(x)$, which is
\begin{eqnarray}\label{contradicting-C}
{\cal C}\nu_{L}(x){\cal C}^{\dagger}=\frac{(1-\gamma_{5})}{2}{\cal C}\nu_{L}(x){\cal C}^{\dagger}=\frac{(1-\gamma_{5})}{2}C\overline{\nu_{L}(x)}^{T}=0,
\end{eqnarray}
and similarly for $\nu_{R}(x)$.
Moreover, the well-known C- and P-violating weak interaction Lagrangian is written as 
\begin{eqnarray}\label{weak-int}
{\cal L}_{{\rm Weak}}&=&(g/\sqrt{2})\bar{e}_{L}\gamma^{\mu}W^{(-)}_{\mu}(x)\nu_{L}+ h.c.\nonumber\\
&=&(g/\sqrt{2})\bar{e}_{L}\gamma^{\mu}W^{(-)}_{\mu}(x)[(1-\gamma_{5})/2]\nu_{L}+ h.c. 
\end{eqnarray}
If one assumes again ${\cal C}\nu_{L}(x){\cal C}^{\dagger}= C\overline{\nu_{L}(x)}^{T}$ as C-transformation law, one obtains ambiguous results, namely, the first expression in \eqref{weak-int} implies that ${\cal L}_{{\rm Weak}}$ is invariant under C, while the second expression implies  ${\cal L}_{{\rm Weak}}\rightarrow 0$~\cite{FT}.
One may find further puzzling aspects arising from the ansatz \eqref{naive-operator-C}.  For example,  one encounters a similar ambiguity in the free action for a Weyl fermion
\begin{eqnarray}\label{free-Weyl}
{\cal L}_{\rm Weyl}&=&\overline{\psi_{L}(x)}i\delslash \psi_{L}(x)\nonumber\\
&=&\overline{\psi_{L}(x)}i\delslash [(1-\gamma_{5})/2]\psi_{L}(x),
\end{eqnarray}
if one uses the classical  transformation rule of charge conjugation $(\psi_{L}(x))^{C}= C\overline{\psi_{L}(x)}^{T}$; the first expression is charge conjugation invariant, while the second expression leads to a vanishing Lagrangian.  We emphasize that those puzzling aspects in \eqref{weak-int} and \eqref{free-Weyl} arise from the assumed classical transformation rule of charge conjugation, $(\nu_{L}(x))^{C}= C\overline{\nu_{L}(x)}^{T}$, irrespective of the presence or absence of the operator {\cal C}. (One may recall that, in Lagrangian field theory,  we first define the classical symmetry operation and then look for  the quantum operator to realize it by Noether theorem or other methods. Those examples,  \eqref{weak-int} and \eqref{free-Weyl}, show that we find no sensible classical operation.)
The ansatz \eqref{naive-operator-C} does not work; the quantity $(\nu_{L}(x))^{C}= C\overline{\nu_{L}(x)}^{T}$ represents a convenient auxiliary object, but not a charge conjugation of $\nu_{L}(x)$. 

In comparison, adopting an alternative convention of charge conjugation (again on an unspecified vacuum), i.e.
\begin{eqnarray}\label{CC_convention}
{\cal C}\nu_{L}(x){\cal C}^{\dagger}= C\overline{\nu_{R}}^T, \ \ \ {\cal C}\nu_{R}(x){\cal C}^{\dagger}= C\overline{\nu_{L}}^T,
\end{eqnarray}
which is suggested by our classical convention for charge conjugation, $\nu_{L}^{c}= C\overline{\nu_{R}}^T$ and $\nu_{R}^{c}= C\overline{\nu_{L}}^T$,  and the charge conjugation of a Dirac field ${\cal C}\nu(x){\cal C}^{\dagger}=C\overline{\nu}^T(x)$, does not lead to any apparently puzzling results. Then we have the explicit form in 
\eqref{Majorana}
\begin{eqnarray}
\psi_{+}(x)&=&\frac{(1+\gamma_{5})}{2}\nu(x)+\frac{(1-\gamma_{5})}{2}\nu^{c}(x),\nonumber\\
\psi_{-}(x)&=&\frac{(1-\gamma_{5})}{2}\nu(x)-\frac{(1+\gamma_{5})}{2}\nu^{c}(x).
\end{eqnarray}
The operator charge conjugation of the right-hand side in the first relation, for example, gives 
\begin{eqnarray}\label{example-conventional-C}
{\cal C}\left[\frac{(1+\gamma_{5})}{2}\nu(x)+\frac{(1-\gamma_{5})}{2}\nu^{c}(x)\right]{\cal C}^{\dagger}&=& \frac{(1+\gamma_{5})}{2}\nu^{c}(x)+\frac{(1-\gamma_{5})}{2}\nu(x)\nonumber\\
&\neq&\frac{(1+\gamma_{5})}{2}\nu(x)+\frac{(1-\gamma_{5})}{2}\nu^{c}(x)
\end{eqnarray}
and the basic requirement of \eqref{operator-C}, ${\cal C}\psi_{+}(x){\cal C}^{\dagger}=\psi_{+}(x)$,  is not satisfied in this case either.
We are therefore unable to maintain the natural operator charge conjugation in \eqref{Majorana} by simply changing the convention.

To resolve these conflicts in a systematic manner, a relativistic analogue of Bogoliubov transformation, $(\nu, \nu^{c})\rightarrow (N, N^{c})$, defined as
\begin{eqnarray}\label{Bogoliubov}
\left(\begin{array}{c}
            N(x)\\
            N^{c}(x)
            \end{array}\right)
&=& \left(\begin{array}{c}
            \cos\theta\, \nu(x)-\gamma_{5}\sin\theta\, \nu^{c}(x)\\
            \cos\theta\, \nu^{c}(x)+\gamma_{5}\sin\theta\, \nu(x)
            \end{array}\right),
\end{eqnarray}
with a suitable parameter $\theta$ has been used for the single flavor seesaw case~\cite{FT, FT2}.  This Bogoliubov transformation (we simply use ``Bogoliubov transformation'' for the more precise ``a relativistic analogue of Bogoliubov transformation'' in the sequel) maps a linear combination of a Dirac field $\nu$ and its charge conjugate $\nu^{c}$ to another Dirac field $N$ and its charge conjugate $N^{c}$; to be precise, all these fields are ``Dirac-type fields''. 
Note that, by definition, $N^{c}=C\bar{N}^{T}$ and $\nu^{c}=C\bar{\nu}^{T}$, and the transformation \eqref{Bogoliubov} satisfies  the (classical) consistency condition $N^{c}=C\bar{N}^{T}$ using the expressions given by the right-hand sides.
We can then show that 
\begin{eqnarray}\label{free-Lagrangian}
{\cal L}&=&\frac{1}{2}\{\bar{N}i\delslash N + \bar{N^{c}}i\delslash N^{c}\}\nonumber\\
&=&\frac{1}{2}\{\bar{\nu}i\delslash \nu + \bar{\nu^{c}}i\delslash \nu^{c}\}
\end{eqnarray}
and the anticommutators are preserved, i.e.,
\begin{eqnarray}\label{anti-comm}
&& \{N(t,\vec{x}), N^{c}(t,\vec{y})\}=\{\nu(t,\vec{x}), \nu^{c}(t,\vec{y})\},\nonumber\\  
&&\{N_{\alpha}(t,\vec{x}), N_{\beta}(t,\vec{y})\}=\{N^{c}_{\alpha}(t,\vec{x}), N^{c}_{\beta}(t,\vec{y})\}=0.
\end{eqnarray}  
Thus, Bogoliubov transformation satisfies the canonicity condition; this anticommutators are established irrespective of the mass values of $\nu(t,\vec{x})$ and $N(t,\vec{x})$. 
We emphasize that the Bogoliubov transformation \eqref{Bogoliubov} preserves the CP symmetry as a unitary operation on quantum fields,
although it does not preserve the transformation properties under $i\gamma^{0}$-parity  and C separately~\cite{FT}.  The present Bogoliubov transformation, which is a canonical transformation, is expected to preserve dynamical properties, but it critically changes the charge conjugation properties.  
A transformation analogous to \eqref{Bogoliubov} has been successfully used in the analysis of neutron-antineutron oscillations~\cite{FT1} and in the discussion of Majorana neutrinos in some classes of seesaw models~\cite{FT}.

The mixing angle $\theta$ in  \eqref{Bogoliubov} is given by the formula 
\begin{eqnarray}\label{mixing-angle}
\sin 2\theta=\frac{\epsilon_{5}/2}{\sqrt{m_{D}^{2}+(\epsilon_{5}/2)^{2}}}
\end{eqnarray}
in the case of a single flavor model~\cite{FT}. If one sets $m_{D}=0$, we have $\theta=\pi/4$ which is independent of $\epsilon_{5}=m_{R}-m_{L}$. Thus the Bogoliubov transformation with $\theta=\pi/4$ is universally valid for a multi-generation model independently of $m_{R}$ and $m_{L}$ as long as the Dirac mass $m_{D}$ is treated as a small perturbation.  
Moreover, for the exact solution \eqref{exact-solution}, it will be shown later that the same $\theta=\pi/4$ is chosen.  We thus adopt $\theta=\pi/4$ for the analysis of {\em three generations} in this paper.

The transformation \eqref{Bogoliubov} with $\theta=\pi/4$ gives
\begin{eqnarray}\label{Bogoliubov-to-Majorana}
\frac{1}{\sqrt{2}}(N(x)+N^{c}(x))&=& \nu_{R}(x)+C\overline{\nu_{R}}^{T}(x),\nonumber\\
\frac{1}{\sqrt{2}}(N(x)-N^{c}(x))&=& \nu_{L}(x)-C\overline{\nu_{L}}^{T}(x),
\end{eqnarray}
namely, two Majorana fermions, 
\begin{eqnarray}\label{Majorana-fermions}
&&\psi_{M}^{(1)}= \frac{1}{\sqrt{2}}(N(x)+N^{c}(x)),\nonumber\\
&&\psi_{M}^{(2)}=\frac{1}{\sqrt{2}}(N(x)-N^{c}(x)),
\end{eqnarray}
are naturally defined in terms of the {\em Bogoliubov quasi-fermion} $N(x)$ in the new vacuum~\footnote{The definition $\psi_{M}^{(2)}=\frac{1}{\sqrt{2}i}(N(x)-N^{c}(x))$ with an imaginary factor $i$ which satisfies $\psi_{M}^{(2)}=C\overline{\psi_{M}^{(2)}}^{T}$ is often used, but this definition requires an anti-unitary ${\cal C}$ to maintain ${\cal C}\psi_{M}^{(2)}{\cal C}^{\dagger}=\psi_{M}^{(2)}$. We stick to the tradition of  unitary ${\cal C}$ with only the time reversal being anti-unitary. In physical applications (with canonical quantization),
our definition of $\psi_{M}^{(2)}=\frac{1}{\sqrt{2}}(N(x)-N^{c}(x))$ with ${\cal C}\psi_{M}^{(2)}{\cal C}^{\dagger}=C\overline{\psi_{M}^{(2)}}^{T}=-\psi_{M}^{(2)}$does not lead to any contradiction. }, with the property:
\begin{eqnarray}
{\cal C}_N|0\rangle_N=|0\rangle_N,
\end{eqnarray}
where the ${\cal C}$ operator in $N$-vacuum is denoted as ${\cal C}_{N}$.  At this moment, we postulate the existence of ${\cal C}_N$ and $|0\rangle_N$ with ${\cal C}_N N(x){\cal C}^{\dagger}_N=C\overline{N}^T(x)$.
Similarly, we postulate the existence of  ${\cal C}_\nu$ and $|0\rangle_\nu$ which satisfy 
\begin{eqnarray}
{\cal C}_\nu|0\rangle_\nu=|0\rangle_\nu,
\end{eqnarray}
and ${\cal C}_\nu \nu(x){\cal C}^{\dagger}_\nu=C\overline{\nu}^T(x)$.
We later discuss if we can justify those postulates in the present model of the seesaw mechanism.

The Majorana fermions satisfy ${\cal C}_{N}\psi_{M}^{(1)}{\cal C}_{N}^{\dagger}=\psi_{M}^{(1)}$ and $\psi_{M}^{(1)}=C\overline{\psi_{M}^{(1)}}^{T}$, and ${\cal C}_{N}\psi_{M}^{(2)}{\cal C}_{N}^{\dagger}= -\psi_{M}^{(2)}$ and $\psi_{M}^{(2)}=-C\overline{\psi_{M}^{(2)}}^{T}$, the first being even and the second odd eigenfield of the charge conjugation operator ${\cal C}_{N}$.
The fields $\psi_{M}^{(1)}$ and $\psi_{M}^{(2)}$  correspond to the conventional definitions of Majorana fermions in terms of Weyl fermions on the right-hand side of \eqref{Bogoliubov-to-Majorana}, but they do not support the operator charge conjugation ${\cal C}_{\nu}$ in the original vacuum 
$|0\rangle_\nu$ as we have demonstrated in \eqref{example-conventional-C}. 

The definition of a Majorana fermion by itself, which is expressed as a linear superposition of fermion and antifermion, implies a certain "condensation" of the fermion number in the vacuum.   (See  Ref.~\cite{chang} for an analysis of the  change of vacua, in a case not directly related to the Bogoliubov transformation.)
The Bogoliubov transformation helps define the eigenstates of charge conjugation ${\cal C}_{N}$ in the new vacuum in a consistent manner,
and the conflicts related to Majorana fermions are resolved in the new vacuum defined by ${\cal C}_{N}$ after the Bogoliubov transformation, which is precisely what the first relation of \eqref{Bogoliubov-to-Majorana} implies.

We can also solve \eqref{Bogoliubov-to-Majorana} in terms of the Majorana fermions \eqref{Majorana-fermions} as 
\begin{eqnarray}\label{Majorana-to-Dirac}
\left(\begin{array}{c}
            \nu(x)\\
            \nu^{c}(x)
            \end{array}\right)
&=& \left(\begin{array}{c}
            (\frac{1+\gamma_{5}}{2})\psi_{M}^{(1)}(x)+(\frac{1-\gamma_{5}}{2})     
            \psi_{M}^{(2)}(x)\\
            (\frac{1-\gamma_{5}}{2})\psi_{M}^{(1)}(x)-(\frac{1+\gamma_{5}}{2})     
            \psi_{M}^{(2)}(x)            
            \end{array}\right).
\end{eqnarray}
The Majorana fermions $\psi_{M}^{(1)}(x)$ and $\psi_{M}^{(2)}(x)$, which agree with $\psi_{+}(x)$ and $\psi_{+}(x)$ in \eqref{6}, respectively, belong to definite representations of the basic symmetries P and T and thus C due to the CPT symmetry of field theory on the Minkowski space-time, if they are chosen as the primary dynamical degrees of freedom. With this choice of fundamental fields, the natural quantum realization of the charge conjugation in \eqref{Majorana-to-Dirac} is ${\cal C}_N$, under which we have
\begin{eqnarray}
&&{\cal C}_{N}\psi_{M}^{(1)}(x){\cal C}_{N}^{\dagger}\rightarrow \psi_{M}^{(1)}(x), \nonumber\\ 
&& {\cal C}_{N}\psi_{M}^{(2)}(x){\cal C}_{N}^{\dagger}\rightarrow -\psi_{M}^{(2)}(x).
\end{eqnarray}
 However, in the left-hand side of \eqref{Majorana-to-Dirac} this operation does not send $\nu$ to $\nu^c$, which would be expected if the operator charge conjugation were preserved. This shows that 
 ${\cal C}_{N}\neq {\cal C}_{\nu}$.
 See also \eqref{6}. This mismatch is caused by the construction of the Majorana fermions {\em via Weyl fermions}, although the classical consistency condition  $\nu^{c}(x)=C\overline{\nu(x)}^{T}$ is satisfied.

We thus conclude that we have to identify the physical Majorana neutrinos with
those defined in terms of Bogoliubov quasi-fermions $N(x)$ in \eqref{Majorana-fermions} to maintain the consistent operator charge conjugation property. Namely, those Majorana neutrinos are the Bogoliubov quasiparticles defined in a new vacuum. This analysis works for the general conversion from Weyl fermions to Majorana fermions, as long as $\epsilon_{5}=m_{R}-m_{L}\neq 0$, namely, 
\begin{eqnarray}
m_{R}\neq m_{L},
\end{eqnarray}
which causes the C (and $i\gamma^{0}$-parity) breaking in the starting Lagrangian in \eqref{3}. (This criterion is more clearly seen in the case of a single flavor~\cite{FT}.) It is possible to exactly diagonalize the C-violating Lagrangian in terms of Weyl fermions as in \eqref{exact-solution} but impossible to rewrite it in terms of Majorana fermions, which are the precise eigenstates of the charge conjugation, as in \eqref{exact-solution2} in a logically consistent manner. This shows that the appearance of Bogoliubov quasiparticle is generic. 

The Bogoliubov transformation is applied to the so-called Type III seesaw model~\cite{foot} also when one regards the extra self-conjugate neutral fermion as providing a right-handed component~\cite{FT2}. 
To include the dimension 5 operator of Weinberg~\cite{weinberg} in the present analysis, which is closely related to the so-called Type II seesaw model~\cite{ma}, 
one may start with the exact solution for a general set of parameters and consider a suitable limiting case such as $m_{R}\rightarrow \infty$ with $m_{L}$ kept fixed in the very end of the analysis.

For the very special case $\epsilon_{5}=m_{R}-m_{L}=0$, the original Lagrangian in \eqref{3} is C and CP invariant and it is written as
\begin{eqnarray}
{\cal L}&=&\frac{1}{2}\{\overline{\psi_{+}}(x)[i\delslash - (m_D+\frac{\epsilon_{1}}{2})]\psi_{+}(x)+ \overline{\psi_{-}}(x)[i\delslash - (m_D-\frac{\epsilon_{1}}{2})]\psi_{-}(x)\}
\end{eqnarray}
in terms of two well-defined Majorana fermions
\begin{eqnarray}
\psi_{\pm}(x)=\frac{\nu(x) \pm \nu^{c}(x)}{\sqrt{2}}.
\end{eqnarray}
In other words, the Majorana fermions are consistently defined in terms of the starting fields $\nu(x)$ and $\nu^{c}(x)$ without the help of the Bogoliubov transformation.
To our knowledge, no practical physical application of this specific case 
is known in neutrino physics.  This case has been discussed in the context of neutron oscillations \cite{chang,
FT1}. 

\section{Exact operator analysis} 

We now want to make the definitions of the vacua $|0\rangle_\nu$ and $|0\rangle_N$ and charge conjugation operators defined on them more explicit.  
 
In the three generation seesaw model \eqref{3}, the exact solution in terms of Weyl fermions is given by \eqref{exact-solution}, namely,
\begin{eqnarray}\label{47x}
{\cal L}
&=&\overline{\nu}(x)i\delslash \nu(x)
-(1/2)[\overline{\nu_{R}}M_{1}\nu_{L}^{c}-\overline{\nu_{R}^{c}}M_{2}\nu_{L}] + h.c.,
\end{eqnarray}
which is written as 
\begin{eqnarray}\label{exact-solution-prime}
{\cal L}
&=&\overline{\nu}(x)i\delslash \nu(x)
+(1/4)[\overline{\nu}(M_{1}+M_{2})\gamma_{5}\nu^{c}-\overline{\nu^{c}}(M_{1}+M_{2})\gamma_{5}\nu]  \nonumber\\
&& -(1/4)[\overline{\nu}(M_{1}-M_{2})\nu^{c}+\overline{\nu^{c}}(M_{1}-M_{2})\nu]
\end{eqnarray}
if one defines the Dirac-type variable
\begin{eqnarray}
\nu(x)=\nu_{L}+\nu_{R}.
\end{eqnarray}
We now define two $3\times 3$ diagonal real matrices
by
\begin{eqnarray}
E_{1}= M_{1}-M_{2}, \ \ \ \ E_{5}= M_{1}+M_{2},
\end{eqnarray}
we then have
\begin{eqnarray}\label{exact-solution-prime}
{\cal L}
&=&\overline{\nu}(x)i\delslash \nu(x)
+(1/4)[\overline{\nu}E_{5}\gamma_{5}\nu^{c}-\overline{\nu^{c}}E_{5}\gamma_{5}\nu]  \nonumber\\
&& -(1/4)[\overline{\nu}E_{1}\nu^{c}+\overline{\nu^{c}}E_{1}\nu].
\end{eqnarray}
Note that the terms with $E_{1}$ are C-invariant while the terms  with $E_{5}$ are C-violating.
When one compares this Lagrangian with the original single flavor Lagrangian \eqref{3} where $m_{D}$, $\epsilon_{1}$ and $\epsilon_{5}$ are set to be  real numbers instead of $3\times 3$ matrices, one recognizes that the exact solution is three copies of the single flavor model with {\em vanishing Dirac mass}.

From the definition of the parameter $\theta$ of the Bogoliubov transformation with $m_{D}=0$ in  \eqref{mixing-angle}, one obtains $\theta=\pi/4$.  The basic Bogoliubov transformation is thus given by 
\begin{eqnarray}
\left(\begin{array}{c}
            N(x)\\
            N^{c}(x)
            \end{array}\right)
&=& \left(\begin{array}{c}
            \frac{1}{\sqrt{2}}[\nu(x)-\gamma_{5}\nu^{c}(x)]\\
            \frac{1}{\sqrt{2}}[\nu^{c}(x)+\gamma_{5}\nu(x)]
            \end{array}\right),
\end{eqnarray}
and one obtains the Lagrangian for the Bogoliubov quasiparticle as 
 \begin{eqnarray}\label{exact-solution2-prime}
 {\cal L}
&=&\overline{N}(x)i\delslash N(x)
-\frac{1}{2}\overline{N}(x)E_{5}N(x)
-\frac{1}{4}[\overline{N}(x)E_{1}N^{c}(x)+\overline{N^{c}}(x)E_{1}N(x)],        
\end{eqnarray}
which is charge conjugation invariant.
The essence of the present Bogoliubov transformation is a CP-preserving canonical transformation which modifies the charge conjugation properties.  One sees that C-noninvariant fermion number violating ``condensate'' with $E_{5}$ in \eqref{exact-solution-prime} is converted to a Dirac mass of the Bogoliubov quasiparticle $N(x)$ in \eqref{exact-solution2-prime}.  $E_{5}$ is an analogue of the ``energy gap'' in the Bardeen--Cooper--Schrieffer (BCS) theory. See also the mass generation in the Nambu--Jona-Lasinio model \cite{nambu}.

Using the Majorana-type fields 
\begin{eqnarray}\label{exact-Majorana}
\psi_{+}(x)&=&\frac{1}{\sqrt{2}}[N(x)+N^{c}(x)],\nonumber\\
\psi_{-}(x)&=&\frac{1}{\sqrt{2}}[N(x)-N^{c}(x)],
\end{eqnarray}
which satisfy the charge conjugation properties $\psi_{+}(x)=C\overline{\psi_{+}}^{T}(x)$ and $\psi_{-}(x)=-C\overline{\psi_{-}}^{T}(x)$ if one defines charge conjugation by $N(x)\rightarrow N^{c}(x)$,
we have
\begin{eqnarray}\label{exact-solution3}
 {\cal L}
&=&\frac{1}{2}\overline{\psi_{+}}(x)[i\delslash - \frac{1}{2} (E_{5}+E_{1})]\psi_{+}(x)
+\frac{1}{2}\overline{\psi_{-}}(x)[i\delslash - \frac{1}{2} (E_{5}-E_{1})]\psi_{-}(x).            
\end{eqnarray}
To define the vacuum and charge conjugation operator in the present model, we assume that  $\psi_{\pm}(x)$ are genuine Majorana fields and examine their consistency with the C-transformation properties of other fermions.
 Since $(E_{5}\pm E_{1})/2$ are diagonal, we have six free Majorana fermions, for which we define the vacuum $|0\rangle_{M} $ in the standard manner,
\begin{eqnarray}
\psi^{(+)}_{\pm}(x)|0\rangle_{M}=0,   
\end{eqnarray}
where $\psi^{(+)}_{\pm}(x)$ stand for the positive frequency components.
It is also straightforward to define the charge conjugation operator for the free fermions, which satisfies
\begin{eqnarray}
{\cal C}_{M}\psi_{+}(x) {\cal C}^{\dagger}_{M}=C\overline{\psi_{+}(x)}^{T}=\psi_{+}(x),
\ \ \ \  {\cal C}_{M}\psi_{-}(x) {\cal C}^{\dagger}_{M}=C\overline{\psi_{-}(x)}^{T}=-\psi_{-}(x), 
\end{eqnarray}
with ${\cal C}_{M}|0\rangle_{M}=|0\rangle_{M}$, following the procedure in the textbook \cite{bjorken}; in fact, the operator charge conjugation has the form ${\cal C}_{M}=\exp[i\pi \hat{n}_{\psi_{-}}]$, with the number operator $\hat{n}_{\psi_{-}}=\sum_{\vec{p},s}a^{\dagger}_{\psi_{-}}a_{\psi_{-}}$ of $\psi_{-}(x)$, and thus acts on $\psi_{+}(x)$ in a trivial manner.

Since we can invert \eqref{exact-Majorana} as 
\begin{eqnarray}
N(x)&=&\frac{1}{\sqrt{2}}[\psi_{+}(x)+\psi_{-}(x)],\nonumber\\
N^{c}(x)&=&\frac{1}{\sqrt{2}}[\psi_{+}(x)-\psi_{-}(x)],
\end{eqnarray}
which satisfy  ${\cal C}_{M}N(x) {\cal C}^{\dagger}_{M}=N^{c}(x)$, one can choose
\begin{eqnarray}
|0\rangle_{N}=|0\rangle_{M}, \ \ \   {\cal C}_{N}={\cal C}_{M}.
\end{eqnarray}
In contrast, we have from \eqref{6} in the notation of the present section,
\begin{eqnarray}
            \left(\begin{array}{c}
            \nu(x)\\
            \nu^{c}(x)
            \end{array}\right)
            &=&\left(\begin{array}{c}
            \frac{1+\gamma_{5}}{2}\psi_{+}(x)+\frac{1-\gamma_{5}}{2}      
            \psi_{-}(x)\\
            \frac{1-\gamma_{5}}{2}\psi_{+}(x)-\frac{1+\gamma_{5}}{2}      
            \psi_{-}(x)            
            \end{array}\right),
\end{eqnarray}
which shows 
\begin{eqnarray}
{\cal C}_{M}\nu(x) {\cal C}^{\dagger}_{M}
&=&{\cal C}_{M}\left[\frac{1+\gamma_{5}}{2}\psi_{+}(x)+\frac{1-\gamma_{5}}{2}\psi_{-}(x)\right] {\cal C}^{\dagger}_{M}\nonumber\\
&=&\left[\frac{1+\gamma_{5}}{2}\psi_{+}(x)-\frac{1-\gamma_{5}}{2}\psi_{-}(x)\right]\nonumber\\
&\neq& \nu^{c}(x).
\end{eqnarray}
We thus conclude ${\cal C}_{\nu}\neq {\cal C}_{M}$, if one defines ${\cal C}_{\nu}\nu(x){\cal C}^{\dagger}_{\nu}= \nu^{c}(x)$. This ${\cal C}_{\nu}(t)$ is time dependent since the C-symmetry thus defined is not a symmetry of the original Lagrangian \eqref{47x}. If one should define the vacuum by ${\cal C}_{\nu}(0)|0\rangle_\nu=|0\rangle_\nu$, then $|0\rangle_M \neq |0\rangle_\nu$.  This shows that {\em the vacuum of Majorana fermions is different from the vacuum of Weyl fermions~\cite{FT2}. } The Majorana fermion in the present context may be properly called a Bogoliubov quasiparticle.

\section{Constraint on the parameters of seesaw models}

The seesaw formula of the small neutrino masses is defined in terms of fermion masses, such as $m_{D}$ and $m_{R}$ in 
\eqref{3}. As is well-known, the fermion masses are linearly divergent in renormalizable field theory and renormalized multiplicatively, such that a small bare mass implies a small renormalized mass. It is thus generally believed that the seesaw formula is stable under renormalization and valid for an arbitrary choice of fermion mass parameters.  

In the course of the analysis of the Bogoliubov transformation, we recognized that the exact formula of the neutrino mass is given by a difference of large masses 
\begin{eqnarray}\label{mass-formula}
m_{\nu}=\sqrt{\left(\frac{m_{R}}{2}\right)^{2}+m_{D}^{2}}-\frac{m_{R}}{2}\simeq 
\frac{m_{D}^{2}}{m_{R}}
\end{eqnarray}
and the famous ratio of two numbers appears as an approximation~\cite{FT}.  This property also holds in the three generation model \eqref{exact-solution3}
\begin{eqnarray}
m_{\nu}=\frac{1}{2}(E_{5}-E_{1}).
\end{eqnarray}   
In the spirit of the analyses in~\cite{weinberg2,susskind}, one may naively expect some kind of naturalness issue such as 
\begin{eqnarray}\label{mass-formula}
\frac{m_{\nu}}{m_{R}}\simeq 
\frac{m_{D}^{2}}{m^{2}_{R}}
\end{eqnarray}
and thus some constraints on possible seesaw models.  As was emphasized above, the fermion mass renormalization is quite stable and thus this formula alone does not give any constraint.
 
We discussed the possible implications of the above relation by making the extra assumptions~\cite{FT}: Technically, we assumed the generic nature of the dimensional regularization (pole subtraction), which eliminates quadratic divergences but preserves hierarchy issues~\cite{fujikawa}. Physically, we assumed the idea of the Coleman--Weinberg mechanism~\cite{coleman}, which determines both the fermion and boson masses in terms of the stationary point of the renormalized effective potential.

For example, if one assumes $m_{D}\sim v$ where $v$ is the vacuum value of the ordinary Higgs boson in \eqref{3}, one obtains the induced Higgs mass by a one-loop fermion correction
\begin{eqnarray}\label{hierarchy}
\Delta m^{2}_{H}\sim \frac{m^{2}_{D}}{v^{2}}m^{2}_{R} \sim  m^{2}_{R}
\end{eqnarray} 
in the dimensional regularization. In the spirit of Coleman and Weinberg, this extra correction in the effective potential induced by the assumed presence of $m_{R}$ needs to be subtracted to maintain the stationary point of the renormalized effective potential at the conventional Higgs vacuum value $v$.  An extra subtraction of the order, in symbolic notation, $m^{2}_{R}- m^{2}_{R}=v^{2}$ is required, or the fine tuning 
\begin{eqnarray}\label{mass-formula}
\frac{v^{2}}{m^{2}_{R}}\simeq \frac{m_{\nu}}{m_{R}}\sim 10^{-26},
\end{eqnarray}
using $m_{\nu}=10^{-2} \ \rm{eV}$ and $m_{R}\sim 10^{15} \ \rm{GeV}$ in the present case.  This shows that arbitrary mass parameters in \eqref{3} are not always natural when analyzed using the above assumptions.

One may define the hierarchy issue as a drastic modification of the mass scale of the Standard Model, which is determined by the vacuum value $v$ of the Higgs field, by higher order quantum corrections.  The above correction \eqref{hierarchy} is then regarded as an example of hierarchy issue induced by $m_{R}$.  When one understands the hierarchy issue in this broad sense, one can show that some parameter regions of specific seesaw models of 
type II~\cite{ma} and III~\cite{foot} also become unnatural by an analysis of the quantum corrections to the Higgs mass term in the renormalized effective potential using the dimensional regularization.

If one accepts the dimensional regularization and the idea of Coleman and Weinberg, one thus obtains a criterion of 
``natural'' seesaw models; if the hierarchy in the bosonic sector is drastically distorted by the extra particles introduced to explain the small neutrino masses, such models are not generally natural.  The models in~\cite{gouvea} and \cite{shaposhnikov}, for example, belong to those natural models consistent with our criterion. So far our analysis  is based on the assumption of the absence of SUSY.   If one allows SUSY, there is an argument that suitable SUSY if discovered below the GUT mass scale may cure all those hierarchy issues~\cite{luis}.

\section{Discussion}

We have identified the Majorana neutrinos as  Bogoliubov quasiparticles  in a natural manner by extending the analysis of a relativistic analogue of the Bogoliubov transformation to the three generations of neutrinos, when C and P are violated with $m_{R}\neq m_{L}$.   The interpretation of the Majorana neutrino as a Bogoliubov quasiparticle is thus generic and it supports our suggestion that the Majorana neutrino should be identified as the first Bogoliubov quasiparticle among elementary particles~\cite{FT2}. The present analysis of the Majorana neutrino as the Bogoliubov quasiparticle may be compared to the recent interest in Majorana fermions in condensed matter physics where the Bogoliubov quasiparticle is well-known, but the idea of Majorana fermions is new~\cite{jackiw, beenakker, wilczek}.

In the course of the analysis of the Bogoliubov transformation, we also mentioned some novel constraints on the parameter regions of seesaw models using the dimensional regularization and the idea of Coleman and Weinberg.
\\

{\it Note added:}

Our relativistic analogue of the Bogoliubov transformation, which was introduced to define the proper charge conjugation
of the Majorana neutrino starting with the C-violating seesaw Lagrangian,  is related to the Pauli--G\"ursey transformation \cite{pauli,gursey}.  We thank M. Fukugita and T. Yanagida for calling the Pauli--G\"ursey transformation to our attention.

\section*{Acknowledgments}
We thank Masud Chaichian for stimulating discussions. This work is supported in part by the Magnus Ehrnrooth Foundation. The support of the Academy of Finland under the Projects no. 136539 and 272919 is gratefully acknowledged.

\end{document}